\documentclass[aps,pra,print,showpacs,superscriptaddress,twocolumn,amsmath,amssymb]{revtex4-1}%double-spaced,
\usepackage{amsfonts}
\usepackage{txfonts}
\usepackage{amsmath}
\usepackage[colorlinks,breaklinks,linkcolor=blue,anchorcolor=blue,citecolor=blue,urlcolor=blue]{hyperref}
\usepackage{graphicx}% Include figure files
\usepackage{dcolumn}% Align table columns on decimal point
\usepackage{bm}% bold math
\usepackage{amssymb}
\usepackage{mathrsfs}
\usepackage[sort&compress]{natbib}
\usepackage{bm,hyperref}
\usepackage{subfigure}
\usepackage{lipsum}
\usepackage{soul}
\usepackage{color}
\usepackage{epstopdf}
\begin{document}

\title{Effect of quantum jumps on non-Hermitian system}
\author{Xiangyu Niu}
\affiliation{Center for Quantum Sciences, School of Physics, Northeast Normal University, Changchun 130024, China.}
\author{Jianning Li}
\affiliation{Center for Quantum Sciences, School of Physics, Northeast Normal University, Changchun 130024, China.}
\author{S. L.  Wu}
\affiliation{School of Physics and Materials Engineering, Dalian Nationalities University, Dalian 116600, China.}
\author{X. X.  Yi}
\email{yixx@nenu.edu.cn}
\affiliation{Center for Quantum Sciences, School of Physics, Northeast Normal University, Changchun 130024, China.}
\date{\today}

\begin{abstract}
One among the possible realizations of non-Hermitian systems is based on open quantum systems by omitting  quantum
jumping terms in the master equation. This is a good approximation at short times where the effects of quantum jumps can be ignored. However, the jumps can  affect the long time dynamics of the system, motivating us to take the jumps into account in these studies. In this paper, by treating the quantum jumps as perturbations, we examine the effect of the quantum jumps on the non-Hermitian system.
For this purpose, we first derive an effective Hamiltonian to describe the dynamics of the open
quantum system based on the master equation, then expand the eigenstates and eigenenergies up to the first and second order in the quantum jumps. Finally, we apply our theory
to a dissipative two-level system and dissipative fermionic superfluids. The effect of quantum jump on the dynamics and the nonequilibrium phase transition is demonstrated and discussed.
\end{abstract}

\maketitle

%\tableofcontents

\section{\label{sec:level1}INTRODUCTION}
In recent years, non-Hermitian (NH) systems \cite{bender2007making} have attracted much attention \cite{ashida2020nonhermitian} from both sides of theoretical and experimental studies. Without the restriction of Hermiticity, non-Hermitian Hamiltonians have been applied to re-examine
well-known quantum systems ranging from single-particle to many-body systems \cite{lee2014heralded,yoshida2019nonhermitian,xu2020topological,
liu2020nonhermitian,yaoEdgeStatesTopological2018}. Interesting features and novel observations are found, including phase transitions
\cite{lourenco2018kondo}, exception points (EPs) \cite{berry2004physics,hassan2017dynamically,chen2017exceptional,lai2019observation},
quantum skin effect \cite{hagaLiouvillianSkinEffect2021,okuma2020topological,lee2020manybody}, non-Bloch bulk-boundary correspondence
\cite{yaoEdgeStatesTopological2018,wang2020defective}, unidirectional zero reflection \cite{shen2018synthetic,yan2020unidirectional} and so on.

Non-Hermitian systems differ from their Hermitian counterparts in many aspects, such as the non-conservation of probability,
 complex-valued eigen-energies and biorthonormal eigenstates \cite{brody2014biorthogonal}.
In order to obtain an effective NH Hamiltonian, many works suggest using open quantum system \cite{lee2014heralded,yamamoto2019theory,liu2020nonhermitian, xu2020topological,yoshida2019nonhermitian,yang2021exceptional} by neglecting quantum jumps, adding reciprocal terms into Hermitian systems \cite{yaoEdgeStatesTopological2018,lee2020manybody,
gong2018topologicala}, or using parametric-amplifier type interactions \cite{wang2019nonhermitian}.
Among them, the most popular scheme is the open system approach, which neglects the quantum jumps in the Lindblad master equation and is valid at the short time limit defined by the loss rate $1/\gamma$ \cite{yamamoto2019theory,yoshida2019nonhermitian,xu2020topological, durrliebliniger}. It is worth addressing that the validity of ignoring the jumps also depends on initial states of the dynamics.

The quantum jumps are associated with the terms in the quantum master equation that act on
both left and right side of the density matrix. From the viewpoint of measurement, the environment can be treated as a device, which
continuously measures the system  and the quantum jumps cause  an abrupt change in  the state of the system. According to the quantum trajectory theory,
the quantum jumps are the terms responsible for the abrupt stochastic change of the wave function. The quantum jumps can
also result in different properties of the exceptional points (EPs). In fact, for Lindbladians with and without quantum jumps
\cite{minganti2019quantum,chen2021quantum}, the EPs can be remarkably different. Connections between the two types of
EPs are established by introducing a hybrid-Liouvillian superoperator, where ``hybrid'' denotes Liouvillian with different strength of jumping terms, which
is capable to describe the passage from a non-Hermitian Hamiltonian to a true Liouvillian including quantum jumps
\cite{minganti2020hybridliouvillian}.

Generally speaking, an analytical solution to the master equation is difficult to obtain due to the huge size of the Hilbert space. Several stochastic approaches, for instance\textbf{,} Monte-Carlo \cite{molmer1993monte, dalibard1992wavefunction} and quantum-trajectory
\cite{daley2014quantum} are put forward. These approaches apply randomness and statistical laws to simulate the occurrence of quantum jumps, which reduce the complexity from $H_N^2$ to $H_N$ with $H_N$ being the size of Hilbert
space of the Hamiltonian. However, their numerical simulations are time-consuming and lack of analytical results. To this extent, the non-Hermitian Hamiltonian is a convenient approach to describe open systems. However, dropping the quantum jump terms might lead to a wrong result. Therefore, the examination on the validity of neglecting the quantum jump terms is an urgent task.

In this paper, by using the effective Hamiltonian approach \cite{yi2001effective},
we propose a method to approximately solve the master equation. The effective Hamiltonian  approach can transform the Lindblad
master equation into a Schr\"odinger-like equation with an effective Hamiltonian, which describes
the dynamics of a composite system consisting of the  system and an auxiliary system. Thus
the dynamics governed by the master equation is transformed into an evolution of a pure state governed by the effective Hamiltonian, and the pure state can be mapped back to the density matrix of the    system. Here we  develop  the
mapping rule  with a   biorthonormal basis. By combining the effective Hamitonian approach with NH-perturbation theory
\cite{sternheim1972nonhermitian,kato2013perturbation}, we formally derive a higher-order
approximate solution to the master equation, and  illustrate our theory with examples.

This paper is organized as follows. In Sec. \ref{sec:Basis}, we introduce  the effective NH Hamiltonian approach and combine it with
the perturbation theory to derive a solution for the density operator. We first assume that the quantum
jump terms in the master equation are negligible, then treat these terms   as perturbations. In Sec. \ref{sec:TLS} and
Sec. \ref{sec:BCS} we apply our method to  two-level system with decoherence  and a dissipative  BCS (Bardeen-Cooper-Schrieffer)
system \cite{bardeen1957microscopic}. We calculate the approximate density operator, energy and the fidelity of initial state by the present theory. The results are discussed and the effect  of quantum jumps on the BCS state is analyzed. Finally,
we conclude  in Sec. \ref{sec:conclusion} .

\section{\label{sec:Basis}formalism}

The Markov master equation is one of the most fundamental descriptions for open systems in quantum theory
\cite{gardiner2004quantum}, which was derived with the  weak coupling assumption and the Markov approximation. The master
equation is  valid in many circumstances, and the solution to the equation obey the basic rules of quantum mechanics such
as trace preserving, complete positivity and Hermiticity. This is the reason  why a wide range of applications have been  developed
in various fields including quantum state preparation \cite{kraus2008preparation}, excitation transfer in light-harvesting
systems, quantum measurement \cite{walls1985analysis} and quantum computation \cite{verstraete2009quantum}. Due to the complexity of the master equation, many methods have been introduced to approximately solve the equation. For examples, in Ref. \cite{kim1996perturbative} the authors developed a short-time perturbative expansion method, and  Ref. \cite{yi2000perturbative} presented  a perturbation  theory by treating the small-loss as the perturbations. Ref. \cite{li2015perturbative}, decomposes  the  Liouvillian supperoperator into two parts, treating the part of dissipators as  the dominant contribution to the system, while  the other parts of the dissipators were treated as perturbations. Based on this, Ref. \cite{li2016resummation} considered a practical model of damping Jaynes-Cumming lattices, in which the interaction between the resonator mode and the qubit was viewed as a perturbation.  Ref. \cite{albert2018lindbladians} introduced a perturbation theory for a time-dependent Lindbladian master equation  with the help of Dyson expansions and linear response theory.

In the following, we will develop a new approach to solve the master equation based on the
perturbation theory for non-Hermitian
systems, the difference is that the jumping terms in the master equation are treated as the perturbation terms.  Let us start with the master equation in the Lindblad form \cite{breuer2002theory,gardiner2004quantum}
\begin{equation}
\dot{\rho}\!=\!-i\left[H_0,\rho\right]\!+\!\sum_m\!\frac{\kappa_m}{2}\!(2F_m\rho F_m^{\dagger}\!-\!F_m^{\dagger}F_m\rho\!
-\!\rho F_m^{\dagger}F_m),\label{eq:ME}
\end{equation}
where $H_0$ is the free Hamiltonian of the system,   $\kappa_m$ is the decay rate for the $m$-th decay channel, and $F_m$ stands for
the eigenoperator of the system, usually named as Lindblad operators. The reduced density matrix $\rho$ remains completely positive
and trace preserving \cite{lindblad1976generators}. However, these break down when the jumping terms are neglected and an effective
non-Hermitian Hamiltonian $H=H_0-i/2\sum_m\kappa_mF^{\dagger}_mF_m$ is persisted to describe the system.
Suppose $H$ is diagonalizable and the eigenvectors satisfy
\begin{equation}
H\vert r_n\rangle=E_n\vert r_n\rangle, H^{\dagger}\vert l_n\rangle=E_n^*\vert l_n\rangle,\label{eq:bio}
\end{equation}
where $\vert r_n\rangle$ and $\vert l_n\rangle$ are the right and left eigenvectors of $H$. In the following
discussion, we will follow the biorthonormal relation $\langle l_m\vert r_n\rangle=\delta_{mn}$,
and the completeness relation  \cite{brody2014biorthogonal}
 \begin{equation}
    \sum_n\vert r_n\rangle\langle l_n\vert=\sum_n\vert l_n\rangle\langle r_n\vert=I.\label{cr}
\end{equation}

Since the biorthonormal eigenvectors are also complete \cite{brody2014biorthogonal}, we can use them to expand the
density matrix. Following  Ref. \cite{yi2001effective}, we can obtain an effective Hamiltonian as long as  the mapping  between the composite system and the density operator is specified. Here we generalize this theory taking different set of eigenstates as the basis. The details of the generalization  can be found in Appendix \ref{app:effectH}. We should address that once the relation is established, the effective Hamiltonian is unique
\begin{equation}
    \widetilde{H}=H-H^{A*}+i\sum_m\kappa_m F_m F^{A*}_m,\label{eq:Htilde}
\end{equation}
here the superscript $A$ denotes the auxiliary system, whose matrix representation satisfies
\begin{align}\langle r_m\vert O^\dagger\vert l_n\rangle=(^A\langle L_n\vert O^A\vert R_m\rangle^A)^*.\label{eq:relation}
\end{align}

Under the mapping rules, the density operator matrix element is now defined as  $\rho_{mn}=\langle l_m\vert\rho\vert l_n\rangle$
and the  Schr\"odinger-like state $\vert\psi_{\rho}\rangle$ reads
\begin{equation}
\vert\psi_{\rho}\rangle=\sum^N_{mn}\rho_{mn}\vert r_m\rangle\vert R_n\rangle^{A*}\rightarrow\rho=
\sum_{mn}\rho_{mn}\vert r_m\rangle\langle r_n\vert,  \label{eq:mapping}
\end{equation}
 here the elements of $\rho$   is defined in basis  $\{\vert r_n\rangle\}$ as $\rho_{mn}=\langle l_m\vert\rho\vert l_n\rangle$,
which is slightly different from the earlier definition $\rho_{mn}=\langle l_m\vert\rho\vert r_n\rangle$
\cite{brody2014biorthogonal}. The trace of the density matrix shall be taken as $\text{Tr}(\rho)=\sum_n\langle l_n\vert
\rho\vert r_n\rangle$, and the average values of a physical observable $O$ could thus be calculated as $\langle O\rangle=\text{Tr}(\rho O)=\sum_n\langle l_n\vert
\rho O\vert r_n\rangle$. Both expansions are feasible, but the matrix representation is slightly different. Actually, the right and left eigenvectors can be connected via
an invertible matrix $A$, i.e., \cite{zhang2019nonhermitian}
\begin{equation}
\vert r_m\rangle=A\vert m\rangle, \vert l_m\rangle=(A^{-1})^\dagger\vert m\rangle,\label{eq:Ainvertible}
\end{equation}
where  $\{\vert m\rangle\}$ is a set of complete orthognomal
basis (See Appendix \ref{app:Vmatrix} for more details).

The effective Hamiltonian $\widetilde{H}$ in Eq. (\ref{eq:Htilde}) can be regarded as a composite system, whose Hilbert space is hence enlarged from $N$ to $N^2$, and the jumping terms in the master equation now
describe the coupling between the system and the ancilla (see Appendix \ref{app:effectH}).
When $\kappa_m$ is small, the interation term $\widetilde{V}=i\sum_m\kappa_m F_m F^A_m$, can be treated as a perturbation.
Following the perturbation theory \cite{sternheim1972nonhermitian} for non-Hermitian systems \cite{bender1999largeorder,
sticlet2022kubo,sticlet2022kubo,buth2004nonhermitian}, we find the first order correction to the $n$th energy and first order
correction to the $n$-th eigenvector,
\begin{eqnarray}\label{eq:per}
e^{(1)}_n&=&\langle \widetilde{\psi}^{(0)}_n\vert \widetilde{V}\vert \psi^{(0)}_n\rangle,\nonumber\\
\vert \psi_n^{(1)}\rangle&=&\sum_{k\neq n}\frac{\langle \widetilde{\psi}^{(0)}_k\vert \widetilde{V}\vert \psi^{(0)}_n\rangle}{e^{(0)}_n-e^{(0)}_k}
\vert \psi^{(0)}_k\rangle,
\end{eqnarray}
where $\{\vert\psi^{(0)}_n\rangle\}$ and $\{\vert\widetilde{\psi}^{(0)}_n\rangle\}$ are the right and left eigenvectors of the effective Hamiltonian without interaction terms, corresponding to eigen-energy $e^{(0)}_n$. Apparently the eigenvectors are actually a direct product of the basis of the two systems. The corresponding $e^{(0)}_n$ is also easy to calculated, because the two subsystems are independent of each other.

Now we are in a position to discuss the dynamics of the open quantum system. By using Eq. (\ref{eq:psidt}) we can obtain the state  $\vert\psi_\rho(t)\rangle$ at time $t$ with an initial state $\vert{\psi}_{\rho}(0)\rangle$ (given by the corresponding initial density matrix). Straightforward calculations show that
\begin{eqnarray}
\vert{\psi}_{\rho}(t)\rangle&=&e^{-i\widetilde{H}t}\vert{\psi}_{\rho}(0)\rangle
=e^{-i\widetilde{H}t}\sum_n\vert\psi_n\rangle\langle\widetilde{\psi_n}
\vert{\psi}_{\rho}(0)\rangle\nonumber\\
&=&\sum_ne^{-ie_nt}\vert\psi_n\rangle\langle
\widetilde{\psi_n}\vert{\psi}_{\rho}(0)\rangle,
\end{eqnarray}
leading to the state of the system at time $t$,
\begin{equation}
\rho(t)=\sum_ne^{-ie_nt}\langle\widetilde{\psi_n}\vert{\psi}_{\rho}(0)\rangle\rho_n \label{eq:sumrho},
\end{equation}
where  $\{\vert\psi_n\rangle\}$, $\{\vert\widetilde{\psi_n}\rangle\}$ are the exact right and left eigenvectors of $\tilde{H}$ with corresponding eigenvalue $e_n$. $\vert\psi_n\rangle$ can be expanded by the complete basis vector $\{\vert r_i\rangle\vert R_j\rangle^{A*}\}$ and $\rho_n$ can be  obtained by the mapping role in Eq. (\ref{eq:mapping}). Namely,  $\vert\psi_n\rangle=\sum_{ij}d_{ij}\vert r_i\rangle\vert R_j\rangle^{A*}\rightarrow\rho_n=\sum_{ij}d_{ij}\vert r_i\rangle\langle r_j\vert$.

From this decomposition, we can find the decay feature of the system \cite{minganti2019quantum}, since it relates closely to the eigenvalues of the effective Hamiltonian $\widetilde{H}$. In other words, the eigenenergies
characterize the decay rates of different eigenstates.  $\widetilde{H}$ has one zero eigenenergy in general, which
corresponds to the steady state of the master equation. As time evolves,
the coefficient \textbf{$e^{-ie_nt}$} is  vanishing for $e_n\neq 0$, and the system reaches its steady state in the long-time limit.
Besides when $e_n\neq 0$, we must have $\text{Tr}(\rho_n)=0$, whereas $\text{Tr}(\rho_n)=1$
when $e_n=0$. This property  protects the density operator to preserve its trace.

\begin{figure}[htbp]
\centering
\subfigure{
\label{Fig. Liouvillian1}
\includegraphics[width=0.24\textwidth]{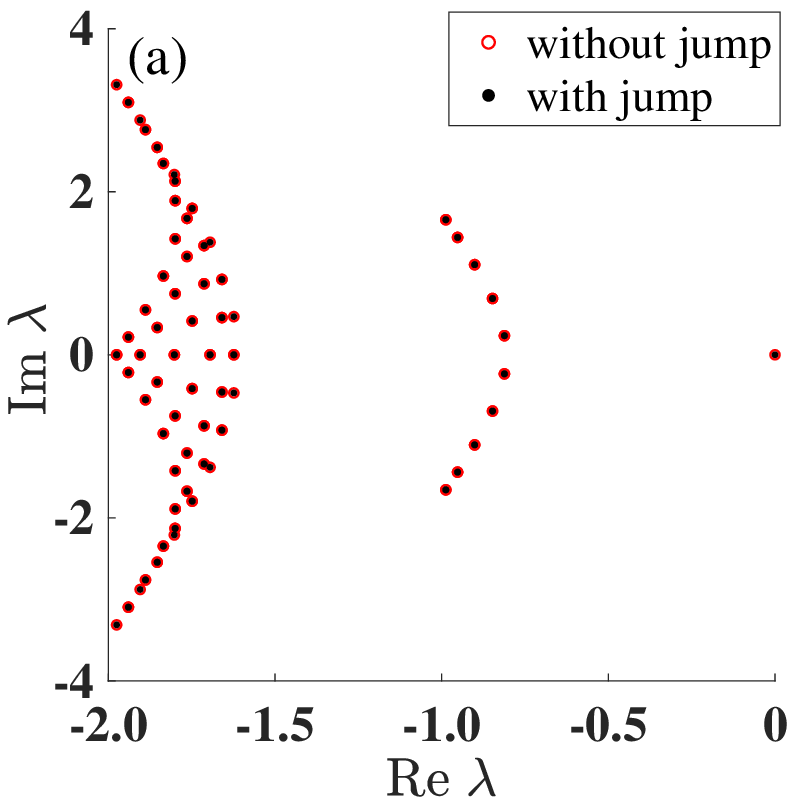}}\subfigure{
\label{Fig. Liouvillian2}
\includegraphics[width=0.24\textwidth]{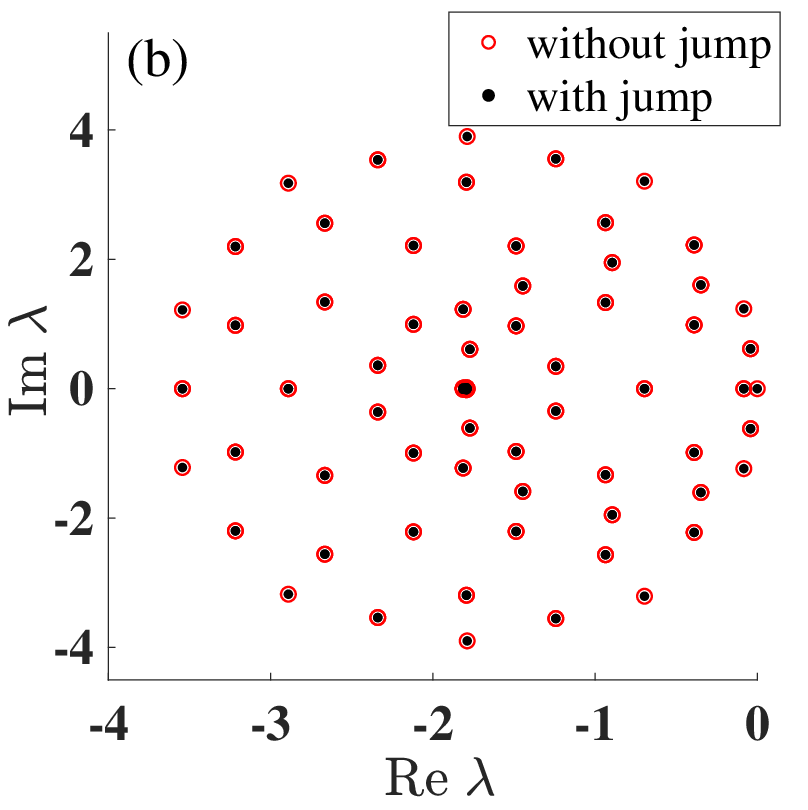}}
\caption{The Liouvillian spectrum of the Hatano-Nelson model for (a) the open boundary condition (OBC) and
(b) the periodic boundary condition (PBC). The red circles represent Liouvillian
without jumps, while the black dots are for  full Liouvillian. We find that they are perfectly overlapping.
Although our numerical calculation is restricted to have two particles at most, the observation holds for more particles. In both figures, the parameters are chosen as $J/\kappa=1$, the number of lattice site is  $n=10$.}
\end{figure}

As aforementioned,  most of the earlier studies  focus on the differences between the spectra of the Liouvillian with and without quantum jumps.
Here we shall emphasize that both the eigenenergies and eigenvectors are important for the dynamics.
Take the model in \cite{gong2018topologicala} as an example, where the authors proposed an implementation scheme in optical lattices for the asymmetric
hopping Hatano-Helson model.
The free Hamiltonian can be written as $H_0=-J\sum_j(c_{j+1}^{\dagger}c_j+c_j^\dagger c_{j+1})$,  where $J$ is
the hopping strength of the lattice and $c_j$ stands for the fermion annihilation operator at site $j$.
When the lattice suffers from the collective one-body loss,  the dynamics is described by a master equation with
a Lindblad operator $F=c_j-ic_{j+1}$ with loss rate $\kappa$. Postselection is used to guarantee that there are no quantum jumps at any time and the system
conserves particle numbers. After neglecting the overall loss, we obtain an effective Hamiltonian $H=\sum_j(J_Rc_{j+1}^{\dagger}
c_j+J_Lc_j^\dagger c_{j+1})$  with the asymmetric hopping strengths  $J_R=-J+\kappa/2,J_L=-J-\kappa/2$. By exact diagonalization \cite{zhang2010exact}, we numerically solve the Liouvillian spectum of the system in both cases
with and without quantum jumps  which are illustrated respectively.   Here, both open
boundary condition (OPC, Fig. \ref{Fig. Liouvillian1}) and period boundary condition (PBC, Fig. \ref{Fig. Liouvillian2}) are considered. From the figures, we find  that the quantum jumps have no effect on the
spectrum of the Liouvillians. In other words, the  Liouvillians with and without quantum jumps have the same spectrum.
Mathematically, this can be understood as that the quantum jumps contribute only to the
block-upper-triangular elements, while the  Liouvillian without quantum jumps is of
block-diagonal form \cite{yoshida2020fate,torres2014closedform,barthel2022superoperator}.

\begin{figure}[htbp]
\centering
\includegraphics[width=0.48\textwidth]{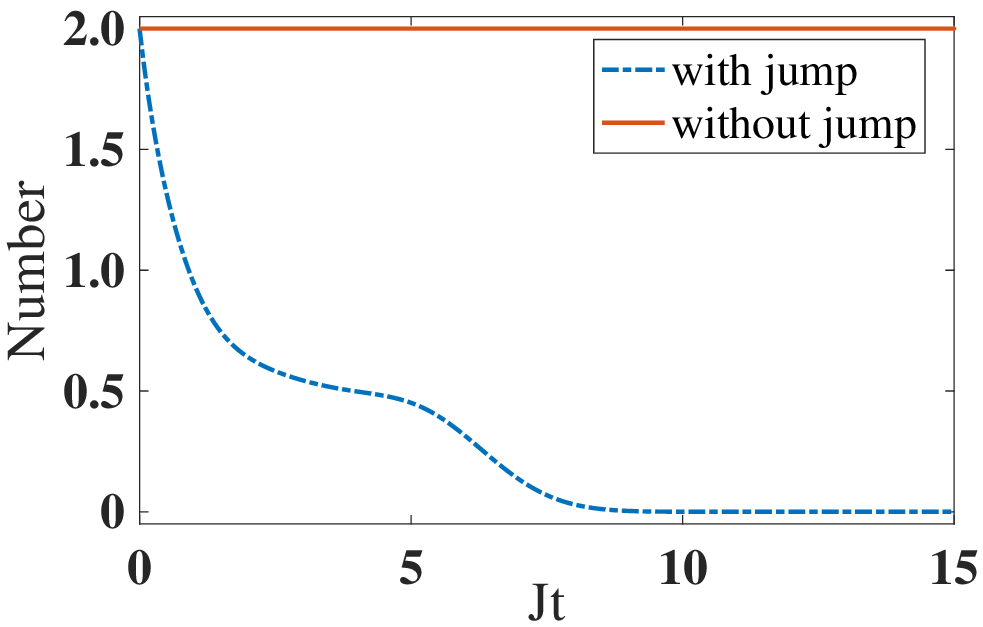}
\caption{The average particle number as a function of time. Blue dash dotted line and  red solid line stand for
the Liouvillians with and without the quantum jumps.} \label{Fig. pnumber}
\end{figure}

Despite the two Liouvillians hold the same spectrum, the dynamics governed by them are totally different.
To be specific, in Fig. \ref{Fig. pnumber} we  show the average particle number $N$ as a function of
time $t$ with initial state $c_1^\dagger c_2^\dagger\vert 0\rangle$ , where $\vert 0\rangle$ is the vacuum  state of fermion. The blue dash dotted line and red line in the figure
are plotted for the system governed by Liouvillians with and
without quantum jumps, respectively. It is obvious that the NH Hamiltonian commutes with the
particle number $\left[H,N\right]=0$, so that the particle number is conserved.
On the other side, the  particle number decreases with time due to the quantum jumps.
From these observations, we find that the same Liouvillian spectrum might leads to different  dynamics
because the eigenstates of the Liouvillians are different.
In the other words, start from an initial state $\rho(0)$ and evolve
 under an non-Hermitian Hamiltonian  $H_{eff}$,  after a tiny time interval $\delta t$, the density matrix  $\rho(\delta t)$ becomes,
\begin{equation}
    \rho(\delta t)=\frac{e^{-iH_{eff}\delta t}\rho(0)e^{iH^{\dagger}_{eff}\delta t}}
    {\text{Tr}(e^{-iH_{eff}\delta t}\rho(0)e^{iH^{\dagger}_{eff}\delta t})}.
\end{equation}
Clearly, the eigenergies and eigenvectors together determine the evolution of the system.

Note that we can also employ the perturbation expansion Eq. (\ref{eq:sumrho})
to calculate  $\rho(\delta t)$ and a normalization is  necessary  because $\rho_n$ in Eq. (\ref{eq:sumrho}) are not
traceless in the present perturbation theory.

Before closing this section, we would like to point out that our scheme is different from approximations in the literature at the following points. The first point  is that we only treat the jumping terms as perturbations---this means that without perturbations the system is governed by a non-Hermitian Hamiltonian, and we focus on whether some of the results and phenomena in the various existing references are sustainable over time. And the second is that the definition of the zero-order steady state may not be so intuitive, because the zero order
equation in general do not satisfies $-i(H\rho_0-\rho_0 H^\dagger)=0$, where $\rho_0=\vert r\rangle\langle l\vert$ and $\{\vert r\rangle\},\{\vert l\rangle\}$ is a set of right and left eigenvectors of the effective non-Hermitian Hamiltonian $H$. In our scheme the perturbation of the energy and eigenvectors of the Hamiltonian in Eq. (\ref{eq:Htilde}) is actually used for the building of evolution equation, as shown in Eq. (\ref{eq:sumrho}).

\section{\label{sec:TLS}application 1: TWO-LEVEL SYSTEM}
In this section,  we illustrate our theory with a dissipative two-level atom. We consider three decoherence channels: including the Bit-Flip and Phase-Flip channels, and their decoherence rates are $\gamma_p$, $\gamma_x$ and $\gamma_z$, respectively.
The dynamics of the system can be described by the following master equation,
\begin{equation}
    \dot{\rho}=-i(H\rho-\rho H^{\dagger})+\gamma_p\sigma_+\rho\sigma_-+\gamma_x\sigma_x\rho\sigma_x+\gamma_z\sigma_z\rho\sigma_z,
\end{equation}
with $H=\frac{\omega}{2}\sigma_z-i\frac{\gamma_p}{2}\sigma_{+}\sigma_{-}
-i\frac{\gamma_x}{2}\sigma_x^2-i\frac{\gamma_z}{2}\sigma_z^2$, where $\sigma_x,\sigma_y,\sigma_z$ are
Pauli matrix, and $\sigma_{\pm}=(\sigma_x\pm i\sigma_y)/2$ are the rising and lowering operators.

Based on the effective Hamiltonian approach, we introduce an ancillary two-level
system with $\sigma^A$ denoting its Pauli matrix. With the basis
spanned by the eigenvectors of $\sigma_z$ and $\sigma_z^A$, with spin up state $\vert 0\rangle$ for the system and
$\vert 0\rangle^A$ for the ancilla, while the spin down states are $\vert 1\rangle$ and $\vert 1\rangle^A$. We can first write out the matrix representation of $H$
\begin{align}
H=\left[
\begin{matrix}
\frac{\omega}{2}-i\frac{\gamma_p}{2}-i\frac{\gamma_x}{2}-i\frac{\gamma_z}{2} & 0\\
0 & -\frac{\omega}{2}-i\frac{\gamma_x}{2}-i\frac{\gamma_z}{2}
\end{matrix}\right],
\end{align}
where the order of basis is $\{\vert 0\rangle,\vert 1\rangle\}$ and they  diagonalize the Hamiltonian $H$. Apparently, $^A\langle 0\vert H^A\vert 0\rangle^A=(\langle 0\vert H^\dagger\vert 0\rangle)^*=\frac{\omega}{2}-i\frac{\gamma_p}{2}-i\frac{\gamma_x}{2}-i\frac{\gamma_z}{2}$, $^A\langle 0\vert H^A\vert 1\rangle^A=(\langle 1\vert H^\dagger\vert 0\rangle)^*={^A}\langle 1\vert H^A\vert 0\rangle^A=(\langle 0\vert H^\dagger\vert 1\rangle)^*=0$ and $^A\langle 1\vert H^A\vert 1\rangle^A=(\langle 1\vert H^\dagger\vert 1\rangle)^*=-\frac{\omega}{2}-i\frac{\gamma_x}{2}-i\frac{\gamma_z}{2}.$ Thus $H^{A*}$ takes
\begin{align}
H^{A*}=\left[
\begin{matrix}
\frac{\omega}{2}+i\frac{\gamma_p}{2}+i\frac{\gamma_x}{2}+i\frac{\gamma_z}{2} & 0\\
0 & -\frac{\omega}{2}+i\frac{\gamma_x}{2}+i\frac{\gamma_z}{2}
\end{matrix}\right],
\end{align}
under its basis $\{\vert 0\rangle^A,\vert 1\rangle^A\}$.

With the above consideration, we can obtain the matrix representation of the free Hamiltonian of the composite system $H-H^{A*}$, which will be treated as the zeroth order Hamiltonian
 $$
H-H^{A*}=\left[
\begin{smallmatrix}
-i(\gamma_p+\gamma_x+\gamma_z) & 0 & 0 & 0\\
0 & \omega-i(\frac{\gamma_p}{2}+\gamma_x+\gamma_z) & 0 & 0\\
0 & 0 & -\omega-i(\frac{\gamma_p}{2}+\gamma_x+\gamma_z) & 0\\
0 & 0 & 0 & -i(\gamma_x+\gamma_z)
\end{smallmatrix}
\right],
$$
the order of basis is $\{\vert 0\rangle\vert 0\rangle^A,\vert 0\rangle\vert 1\rangle^A,\vert 1\rangle\vert 0\rangle^A,\vert 1\rangle\vert 1\rangle^A\}$.
Similarly, the
quantum jumps, i.e., the third terms in Eq. (\ref{eq:Htilde}), which describe the coupling  between the two systems read,
$$
\begin{bmatrix}
i\gamma_z & 0 & 0 & i\gamma_x\\
0 & -i\gamma_z & i\gamma_x & 0\\
0 & i\gamma_x & -i\gamma_z & 0\\
i\gamma_p+i\gamma_x & 0 & 0 & i\gamma_z
\end{bmatrix},
$$
 we will treat this coupling as a perturbation. By the perturbation theory given in Eq. (\ref{eq:per}),
the first and second order corrections to the eigenenergies and  the  right eigen-vectors
can be given by,

\begin{flalign}
\begin{split}
&E^{(0)}_1=-i(\gamma_p+\gamma_x+\gamma_z),E^{(0)}_2=\omega-i(\frac{\gamma_p}{2}+\gamma_x+\gamma_z),\\
    &E^{(0)}_3=-\omega-i(\frac{\gamma_p}{2}+\gamma_x+\gamma_z),E^{(0)}_4=-i(\gamma_x+\gamma_z).\\
    &E^{(1)}_1=i\gamma_z,E^{(1)}_2=-i\gamma_z,E^{(1)}_3=-i\gamma_z,E^{(1)}_4=i\gamma_z.\\
    &E^{(2)}_1=\frac{-i\gamma_x(\gamma_p+\gamma_x)}{\gamma_p},E^{(2)}_2=-\frac{\gamma_x^2}{2\omega},\\
    &E^{(2)}_3=\frac{\gamma_x^2}{2\omega},E^{(2)}_4=\frac{i\gamma_x(\gamma_p+\gamma_x)}{\gamma_p}.\\
    &\vert\psi^{(1)}_1\rangle=-\frac{(\gamma_p+\gamma_x)}{\gamma_p}\vert\psi^{(0)}_4\rangle,\vert\psi^{(1)}_2\rangle=\frac{i\gamma_x}{2\omega}\vert\psi^{(0)}_3\rangle,\\
    &\vert\psi^{(1)}_3\rangle=-\frac{i\gamma_x}{2\omega}\vert\psi^{(0)}_2\rangle,\vert\psi^{(1)}_4\rangle=\frac{\gamma_x}{\gamma_p}\vert\psi^{(0)}_1\rangle.\\
    &\vert\psi^{(2)}_1\rangle=\vert\psi^{(2)}_2\rangle=\vert\psi^{(2)}_3\rangle=\vert\psi^{(2)}_4\rangle=0.
\end{split}
\end{flalign}

\begin{figure}[htbp]
\centering
\includegraphics[width=0.48\textwidth]{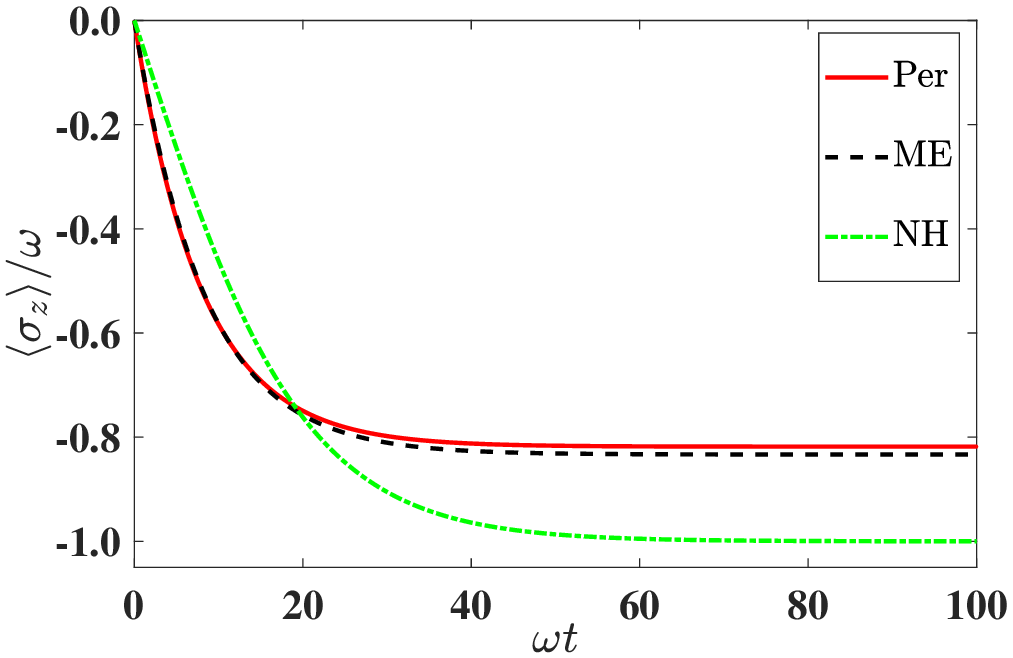}
\caption{The average of $\sigma_z$ at different time $t$. The results are given
by the master equation without any approximation (Per, black dashed line), non-Hermitian Hamiltonian
 (NH, green dash-dotted line), and the perturbation theory (ME, red solid line). The parameters chosen are
 $\gamma_p=0.1\omega,\gamma_x=0.01\omega,\gamma_z=0.5\omega$. }\label{Fig. sub.compare}
\end{figure}

To show the validity of the perturbation theory, we  present  in  Fig. \ref{Fig. sub.compare}
the comparison  between the numerical results given by ME, NH and the perturbation theory.
We find that the NH approximation is close to the numerical result in short time limit,
but it gradually deviates and finally reaches its steady state, which is totally different
from the steady state of the master equation.  The results given by the perturbation theory
are in good agreement with that given by the master equation (or the Liouvillian). Thus we can claim that
the  perturbation theory based on the non-Hermitian Hamiltonian might be a good method to deal with
non-Hermitian systems. Of course this example is easy to solve exactly as the Hilbert space is small. In the next
section, we will present a many-body system to exemplify  the perturbation theory.

\section{\label{sec:BCS} application 2: Effect of quantum jumps on the non-Hermitian BCS states}
In recent years, open many-body systems become active in various  fields from quantum optics to condensed matter. However the master equation of many-body open systems  is   difficult to solve \cite{mcdonald2021nonequilibrium,
yaoEdgeStatesTopological2018}.
 Here  we  apply our perturbation theory to such systems, taking  the NH BCS model as an example \cite{yamamoto2019theory}.

Due to inelastic collisions, the atoms in the BCS system
suffer from two-body loss and  atoms would leave the system with time, such a system
can be described by an effective Markovian master equation  \cite{durrliebliniger}.
In the case that the quantum jumps could be neglected, the earlier study \cite{yamamoto2019theory} showed
that when the interaction strength is not so strong, the superfluid suffer a breakdown and restoration transition occurs as the dissipation increases. Whereas in strong-dissipation limit the superfluid phase
would never be broken. This gives rise to a question: what happens if the quantum jumps can not be neglected?

To answer this question concretely, we consider the 1D model  
in Ref. \cite{yamamoto2019theory}. Here $H_S=\sum_{\emph{\textbf{k}}\sigma}\xi_{\emph{\textbf{k}}}
c^\dagger_{\emph{\textbf{k}}\sigma}c_{\emph{\textbf{k}}\sigma}$ describes the free Hamiltonian of the lattice, where $\xi_{\emph{\textbf{k}}}=\epsilon_{\emph{\textbf{k}}}-\mu$, $\epsilon_{\emph{\textbf{k}}}$ stands for the energy dispersion and $\mu$ is the chemical potential. The interaction Hamiltonian \textbf{$H_I=-U_0\sum_ic^\dagger_{i\uparrow}c_{i\downarrow}^\dagger c_{i\downarrow}c_{i\uparrow}$} (take $\hbar$=1). $c_{i\sigma}$ ($c_{\emph{\textbf{k}}\sigma}$) denote the annihilation operators of a spin-$\sigma\in\{\uparrow,\downarrow\}$ fermion at site $i$ (with momentum $\emph{\textbf{k}}$). Consider the system undergoes inelastic collisions,  the dynamics of the system is governed by
\begin{equation}
\dot{\rho}=-i(H_{\text{eff}}\rho-\rho H_{\text{eff}}^{\dagger})+\kappa\sum_i
L_i\rho L_i^{\dagger},
\end{equation}
where $\kappa$ is the loss rate and $L_i=c_{i\downarrow}c_{i\uparrow}$, and \textbf{$H_{\text{eff}}=\sum_{\emph{\textbf{k}}\sigma}\xi_{\emph{\textbf{k}}}c^\dagger_{\emph{\textbf{k}}\sigma}
c_{\emph{\textbf{k}}\sigma}-\sum_{\emph{\textbf{k}}\emph{\textbf{k}}^{\prime}}U_1/Nc^{\dagger}_{\emph{\textbf{k}}\uparrow}
c^{\dagger}_{-\emph{\textbf{k}}\downarrow}c_{-\emph{\textbf{k}}^{\prime}\downarrow}
c_{\emph{\textbf{k}}^{\prime}\uparrow}$}.
Here $N$ is the number of lattice cites and \textbf{$U_1=U_0+i\kappa/2$} is the complex interaction strength.

\begin{figure}[htbp]
\centering
\subfigure{
\label{Fig:delta01}
\includegraphics[width=0.24\textwidth]{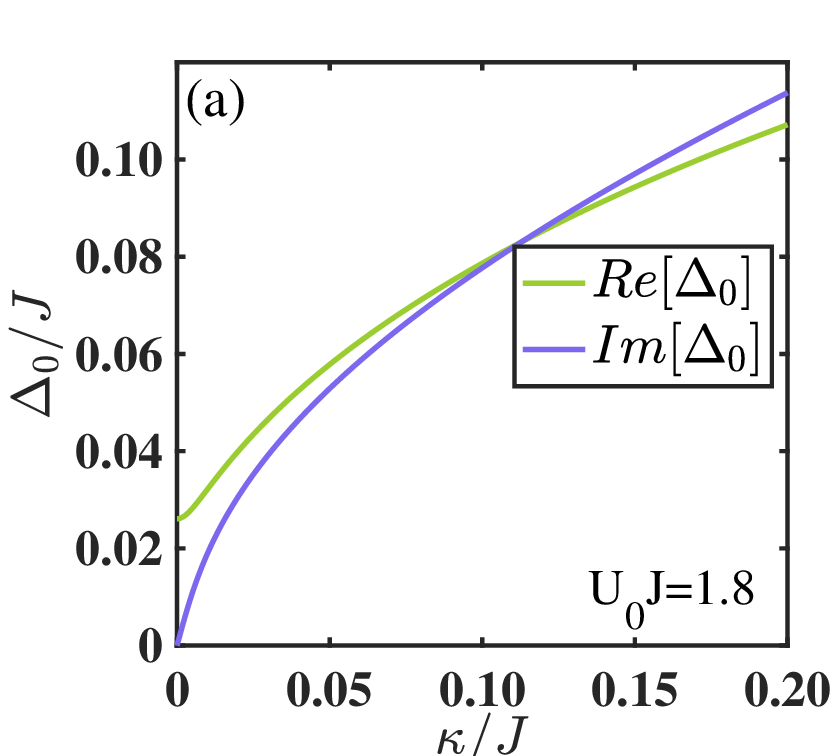}}\subfigure{
\label{Fig:Ec1}
\includegraphics[width=0.24\textwidth]{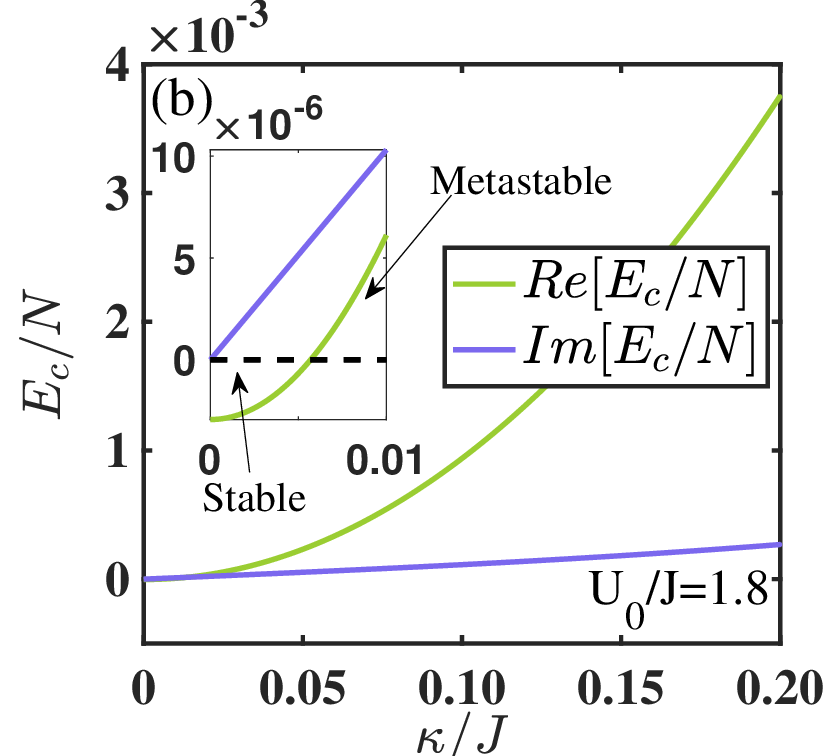}}
\centering
\subfigure{
\label{Fig:delta02}
\includegraphics[width=0.24\textwidth]{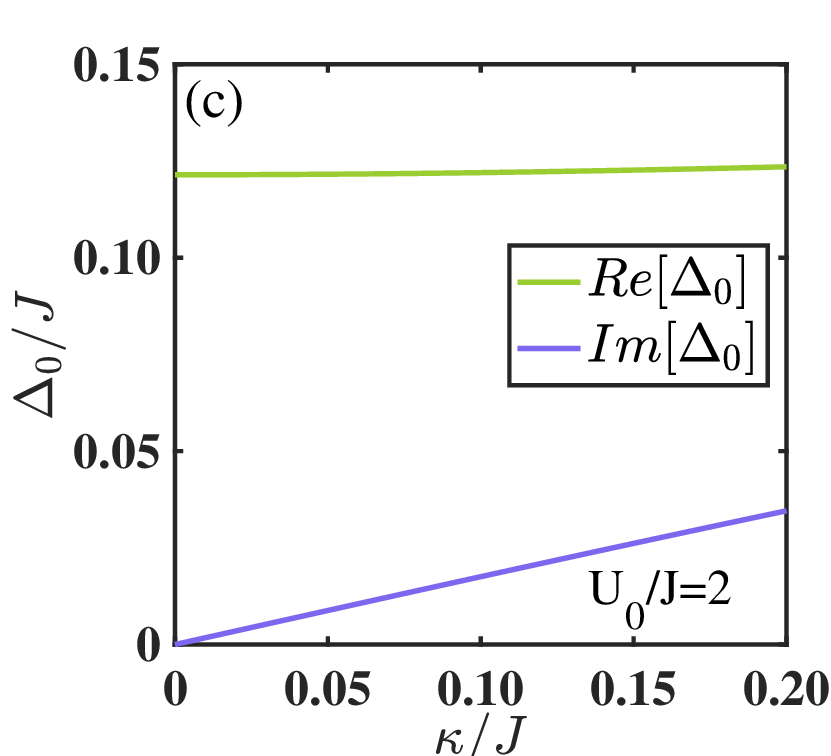}}\subfigure{
\label{Fig:Ec2}
\includegraphics[width=0.24\textwidth]{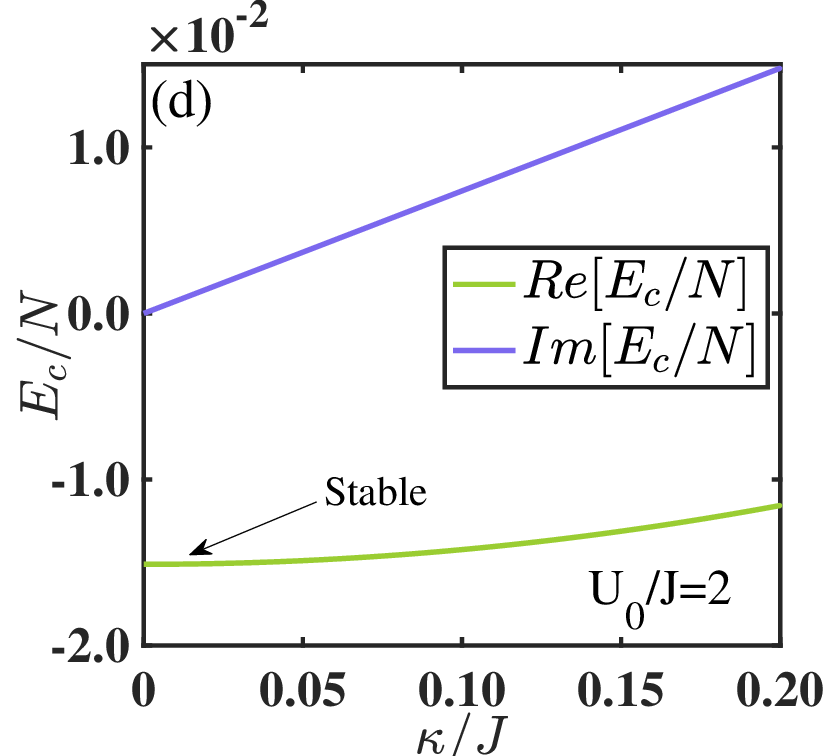}}
\caption{(a)(c) Numerical solution of non-zero superfluid gap $\delta_0$, (b)(d) condensation energy $E_c$, calculated by Eq. (\ref{eq:gap}) and Eq. (\ref{eq:condense}) respectively, as a function that changes with loss rate $\kappa$. Parameters are chosen as $N=10,  \kappa=0.1J$,  $U_0/J=1.8$ for (a)(b) while in (c)(d) $U_0/J=2$. The former system is metastable at most parameters, with stable superfluidity only in the weak dissipation limit. The latter one is always stable for strong attractive interaction $U_0$.
}
\end{figure}

In order to diagonalize the Hamiltonian, the mean-field approximation is applied \cite{yamamoto2019theory},  where the quasiparticles obey
neither Fermi nor Bose statistics, since a NH Hamiltonian cannot be diagonalized by unitary transformations.
Under the mean-field (MF) approximation, Hamiltonian $H_{\text{eff}}$ reduces to
\begin{eqnarray}
H_{\text{MF}}=\sum_{\emph{\textbf{k}}}E_{\emph{\textbf{k}}}
(\bar{\gamma}_{\emph{\textbf{k}}\uparrow}\gamma_{\emph{\textbf{k}}\uparrow}
+\bar{\gamma}_{-\emph{\textbf{k}}\downarrow}\gamma_{-\emph{\textbf{k}}\downarrow})
-\sum_{\emph{\textbf{k}}}E_{\emph{\textbf{k}}}\label{eq:HMF},
\end{eqnarray}
with $E_{\emph{\textbf{k}}}=\sqrt{\xi^2_{\emph{\textbf{k}}}+\Delta_0^2}$, where $\Delta_0=-U_1/N\sum_{\emph{\textbf{k}}}\langle c_{-\emph{\textbf{k}}\downarrow}c_{\emph{\textbf{k}}\uparrow}\rangle$ is
the order parameter (gap function) of the superfluid. In the $\beta\rightarrow0$ limit, the order parameter
can be  established  with NH path integral approach \cite{yamamoto2019theory} or
self-consistency method \cite{fernandeslecture}, which is given by the NH gap equation
\begin{equation}
    \frac{N}{U_1}=\sum_{\emph{\textbf{k}}}\frac{1}{2\sqrt{\xi^2_{\emph{\textbf{k}}}+\Delta_0^2}},\label{eq:gap}
\end{equation}
when $\Delta_0$ takes zero, such phase is denoted as ``normal state'', where the gap equation has only a trivial solution. For most cases $\Delta_0$ is a complex number.

The quasiparticle operators in Eq. (\ref{eq:HMF}) can be written as
\begin{eqnarray}
\bar{\gamma}_{\emph{\textbf{k}}\uparrow}=
u_{\emph{\textbf{k}}}c^{\dagger}_{\emph{\textbf{k}}\uparrow}
-v_{\emph{\textbf{k}}}c_{-\emph{\textbf{k}}\downarrow}, \bar{\gamma}_{-\emph{\textbf{k}}\downarrow}=v_{\emph{\textbf{k}}}
c_{\emph{\textbf{k}}\uparrow}+u_{\emph{\textbf{k}}}
c^{\dagger}_{-\emph{\textbf{k}}\downarrow},\\
\gamma_{\emph{\textbf{k}}\uparrow}=u_{\emph{\textbf{k}}}c_{\emph{\textbf{k}}\uparrow}
-v_{\emph{\textbf{k}}}c^{\dagger}_{-\emph{\textbf{k}}\downarrow}, \gamma_{-\emph{\textbf{k}}\downarrow}=v_{\emph{\textbf{k}}}c^{\dagger}_{\emph{\textbf{k}}\uparrow}
+u_{\emph{\textbf{k}}}c_{-\emph{\textbf{k}}\downarrow},
\end{eqnarray}
with $u_{\emph{\textbf{k}}}=\sqrt{\frac{E_{\emph{\textbf{k}}}
+\xi_{\emph{\textbf{k}}}}{2E_{\emph{\textbf{k}}}}}$, $v_{\emph{\textbf{k}}}=\sqrt{\frac{E_{\emph{\textbf{k}}}
-\xi_{\emph{\textbf{k}}}}{2E_{\emph{\textbf{k}}}}}$.
In the above derivation, the symmetry $H_{\text{MF}}^*=H_{\text{MF}}^{\dagger}$ has been used, which could be found from the matrix representation in terms of Fock states. In fact, under
 such representation, the  non-Hermiticity of the  system attributes to the complex
diagonal elements, which implies that the left eigenvector
$\vert L_n\rangle$ of the Hamiltonian  is exactly the complex conjugation of right one $\vert R_n\rangle$ \cite{yamamoto2019theory}.

\begin{figure}[htbp]
\centering
\subfigure{
\label{Fig. sub.1}
\includegraphics[width=0.24\textwidth]{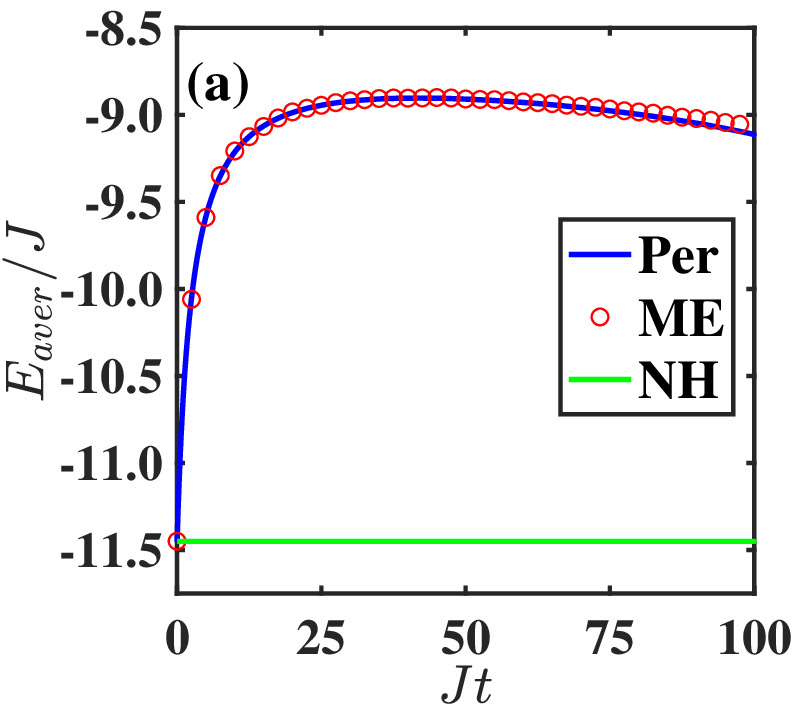}}\subfigure{
\label{Fig. sub.2}
\includegraphics[width=0.24\textwidth]{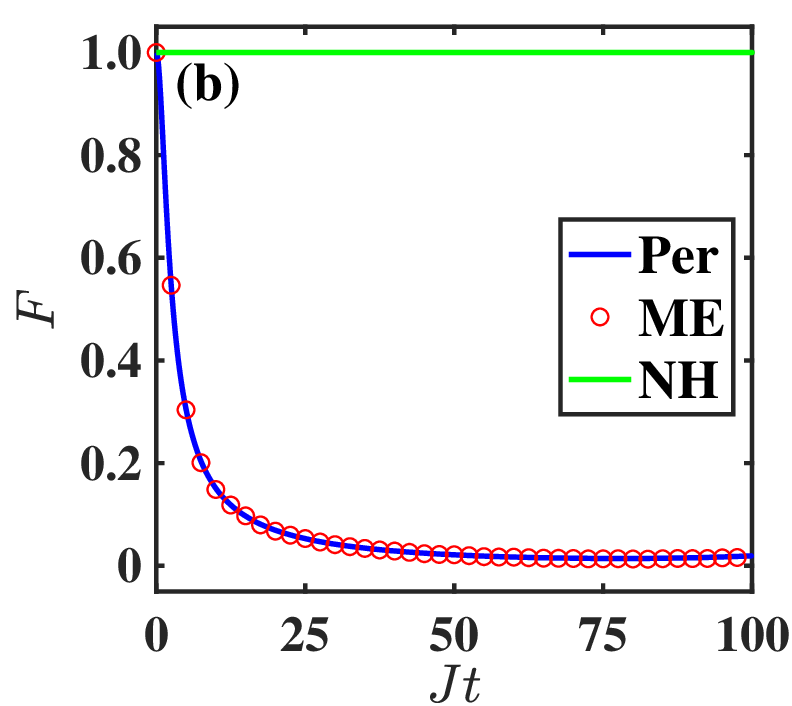}}
\caption{(a) The average energy $E_{aver}=\langle H_S\rangle$ and
 (b) the fidelity of initial state governed by perturbation method (blue solid line), master equation (red circles) and non-Hermitian Hamiltonian (green solid line), respectively. Here we take   $\epsilon_{\emph{\textbf{k}}}=-2J\cos{\emph{\textbf{k}}}$, here $J$ is the hopping amplitude, and the loss rate $\kappa$ is set to be $0.1J$ and $\mu=0$. The interaction strength is set to be $U_0=1.8J$, the number of
 lattice sites $N$ takes 10. The order parameter $\Delta_0/J=0.0786+0.0777i$ (obtained by solving Eq.
(\ref{eq:gap})), and the system is assume to stay in the superfluid state (metastable).
}
\end{figure}

In the following discussion, we would focus on the ground state of the system.
To clarify the discussion, we write down the right and left ground state,
\begin{eqnarray}
\vert r_0\rangle&=&\vert BCS\rangle_R=\prod_{\emph{\textbf{k}}}(u_{\emph{\textbf{k}}}
+v_{\emph{\textbf{k}}}c^{\dagger}_{\emph{\textbf{k}}\uparrow}c^{\dagger}_{-\emph{\textbf{k}}\downarrow})\vert 0\rangle,\nonumber\\
\vert l_0\rangle&=&\vert BCS\rangle_L=\prod_{\emph{\textbf{k}}}(u^*_{\emph{\textbf{k}}}+v^*_{\emph{\textbf{k}}}c^{\dagger}_{\emph{\textbf{k}}\uparrow}c^{\dagger}_{-\emph{\textbf{k}}\downarrow})\vert 0\rangle,
\end{eqnarray}
where $\vert 0\rangle$ is the vacuum state of the fermions.  It is easy to find that $H\bar{\gamma}_{\emph{\textbf{k}}\sigma}\vert BCS\rangle_R=E_{\emph{\textbf{k}}}\bar{\gamma}_{\emph{\textbf{k}}\sigma}\vert BCS\rangle_R,H^{\dagger}{\gamma}^{\dagger}_{\emph{\textbf{k}}\sigma}\vert BCS\rangle_L=E^*_{\emph{\textbf{k}}}{\gamma}^{\dagger}_{\emph{\textbf{k}}\sigma}\vert BCS\rangle_L.$ Here $H$ is Hamiltonian $H_{MF}$ except the constant $-\sum_{\emph{\textbf{k}}}E_{\emph{\textbf{k}}}$ is neglected.
In this way, all left and right eigenvectors $\{\vert l_n\rangle\},\{\vert r_n\rangle\}$ of the effective Hamiltonian $H$
can be constructed.

In Hermitian case, 
the superfluidity of the system arises from the  non-zero gap function, since the energy spectrum will always have a gap in order to excite quasiparticles, even if $\xi_{\emph{\textbf{k}}}$ takes $0$ \cite{fernandeslecture}. However, for NH systems, such defined superfluid may be metastable,
distinguished by the sign of the real part of the condensation energy 
$E_c$
\begin{align}
   E_c=\frac{N}{U_1}\Delta_0^2-\sum_{\emph{\textbf{k}}}(\sqrt{\xi^2_{\emph{\textbf{k}}}+\Delta_0^2}-|\xi_{\emph{\textbf{k}}}|), \label{eq:condense}
\end{align}
actually, $E_c$ represents the difference in the ground-state energy between the superﬂuid and normal states. For positive $\text{Re}(E_c)$, the system is metastable, while a negative $\text{Re}(E_c)$ leads to a stable superﬂuid solution \cite{yamamoto2019theory}. Fig. \ref{Fig:delta01}-\ref{Fig:Ec2} show real and imaginary part of the gap function $\Delta_0$ and the condensation energy $E_c$ when $U_1=1.8J$ and $2J$, respectively. Apparently, the NH gap equations have nontrivial solutions, for small $U_0$, the system is stable only at the small dissipation limit. As attractive interaction strength $U_0$ gets stronger, the system remains stable.

In order to analyse the jump involved circumstance, suppose the system is initially prepared in a NH steady state $\rho_0,$ satisfying  $\dot{\rho}_0=-i(H_{\text{eff}}
\rho_0-\rho_0 H_{\text{eff}}^{\dagger})=0$ and $\text{Tr}(\rho_0)=1$. Clearly we can choose
$\rho_0=\mathcal{N}\vert r_0\rangle\langle r_0\vert$ as an initial state that meets the requirement. Here $\mathcal{N}$ is the normalization coefficient that satisfies $\text{Tr}(\rho_0)=\sum_n\langle l_n\vert\rho_0\vert r_n\rangle=1$. Now we apply the perturbation
theory into the NH BCS system. Consider that our system has only
two-body loss,  we then restrict the Hilbert space enclosing at most 2 quasiparticles. The corresponding eigenergies
are (the constant  in Eq. (\ref{eq:HMF}) is omitted) $\{0,0-2E_{\emph{\textbf{k}}_1}^*,0-2E_{\emph{\textbf{k}}_2}^*\dots,
2E_{\emph{\textbf{k}}_1}-0,2E_{\emph{\textbf{k}}_1}-2E_{\emph{\textbf{k}}_1}^* \dots,\dots,2E_{\emph{\textbf{k}}_n}-0,
2E_{\emph{\textbf{k}}_n}-2E_{\emph{\textbf{k}}_1}^*\dots,
2E_{\emph{\textbf{k}}_n}-2E_{\emph{\textbf{k}}_1}^*\}$. In this Hilbert space, the matrix
representation of the quantum  jumps
$L_{\emph{\textbf{k}}_0}$ takes
\begin{equation}
\begin{bmatrix}
u_{\emph{\textbf{k}}_0}v_{\emph{\textbf{k}}_0} & 0 &\cdots & u_{\emph{\textbf{k}}_0}^2  & \cdots & 0\\
0 & u_{\emph{\textbf{k}}_0}v_{\emph{\textbf{k}}_0} & \cdots  & 0  & \cdots & 0\\
\vdots\ & \vdots  & \ddots & \vdots & \cdots & 0 \\
-v_{\emph{\textbf{k}}_0}^2 & 0 & \cdots & -u_{\emph{\textbf{k}}_0}v_{\emph{\textbf{k}}_0} & \cdots & 0\\
\vdots & \vdots & \vdots & \vdots & \ddots & 0\\
0 & 0 & \cdots & 0 & \dots & u_{\emph{\textbf{k}}_0}v_{\emph{\textbf{k}}_0}\label{eq:matrix}
\end{bmatrix},
\end{equation}\\
 here the order of the basis is arranged as $\{\vert r_0\rangle,\bar{\gamma}_{\emph{\textbf{k}}_1\uparrow} \bar{\gamma}_{-\emph{\textbf{k}}_1\downarrow}\vert r_0\rangle,\\\dots,\bar{\gamma}_{\emph{\textbf{k}}_n
\uparrow}\bar{\gamma}_{-\emph{\textbf{k}}_n\downarrow}\vert r_0\rangle\}\{\langle l_0\vert,\langle l_0\vert\gamma^{\dagger}_{-\emph{\textbf{k}}_1
\downarrow}\gamma^{\dagger}_{\emph{\textbf{k}}_1\uparrow},\dots,\langle l_0\vert\gamma^{\dagger}_{-\emph{\textbf{k}}_n\downarrow}
\gamma^{\dagger}_{\emph{\textbf{k}}_n\uparrow}\}$.
Notice that  the diagonal elements with a minus sign result from the contribution of $\bar{\gamma}_{\emph{\textbf{k}}_0\uparrow}\bar{\gamma}_{-\emph{\textbf{k}}_0\downarrow}\vert r_0\rangle,\langle l_0\vert\gamma^{\dagger}_{-\emph{\textbf{k}}_0\downarrow}\gamma^{\dagger}_{\emph{\textbf{k}}_0\uparrow}$.

By the definition given in Eq. (\ref{eq:Htilde}), the first two terms $H-H^{A*}$ can be taken as zero-order Hamiltonian and the
third term $i\sum_m\kappa_mF_mF^{A*}_m$ is the perturbation. Collecting the results in \cite{yamamoto2019theory}, we find that
the zeroth energy of the ground state is

\begin{equation}
E^{(0)}_0=-2\sum_{\emph{\textbf{k}}}\text{Im}(E_{\emph{\textbf{k}}}),
\end{equation}
by the non-Hermitian perturbation theory, the first order correction to the energy of the  ground state is given by

\begin{equation}
    E^{(1)}_0=-\kappa\sum_{\emph{\textbf{k}}}\vert u_{\emph{\textbf{k}}}\vert^2\vert v_{\emph{\textbf{k}}}\vert^2.
\end{equation}

Here we want to emphasize that the complex constant eigenenergy of the system
can be safely ignored, similar to Hermitian systems  that  constants  can not affect their dynamical features.  This can be interpreted
as a gauge shift $H\rightarrow H+cI$, where $I$ is the identity operator and $c$ is a complex c-number.
In other words, the dynamics under $H$ and $ H+cI$ is the same \cite{zloshchastiev2014comparisona}.
By our theory, the first order corrections
 to the left and right ground states are (unnormalized) \begin{widetext}
\begin{equation}
\begin{aligned}\label{eq:firstorder}
\vert r_0^{(1)}\rangle=\sum_{\emph{\textbf{k}}}\frac{-i\kappa u_{\emph{\textbf{k}}}v_{\emph{\textbf{k}}}v_{\emph{\textbf{k}}}^{*2}}{2E_{\emph{\textbf{k}}}^*}\vert r_0\rangle\gamma^{\dagger}_{{\emph{\textbf{k}}}\uparrow}\gamma^{\dagger}_{-{\emph{\textbf{k}}}\downarrow}\vert l_0\rangle+\sum_{\emph{\textbf{k}}}\frac{i\kappa u_{\emph{\textbf{k}}}^{*}v_{\emph{\textbf{k}}}^{*}v^2_{\emph{\textbf{k}}}}{2E_{\emph{\textbf{k}}}}\bar{\gamma}_{\emph{\textbf{k}}\uparrow}\bar{\gamma}_{-\emph{\textbf{k}}\downarrow}
\vert r_0\rangle\vert l_0\rangle-
\sum_{\emph{\textbf{k}}}\frac{\kappa\vert v_{\emph{\textbf{k}}}\vert^4}{4\text{Im}(E_{\emph{\textbf{k}}})}\bar{\gamma}_{\emph{\textbf{k}}\uparrow}\bar{\gamma}_{-\emph{\textbf{k}}\downarrow}\vert r_0\rangle\gamma^{\dagger}_{{\emph{\textbf{k}}}\uparrow}\gamma^{\dagger}_{-{\emph{\textbf{k}}}\downarrow}\vert l_0\rangle,\\
\vert l^{(1)}_0\rangle=\sum_{\emph{\textbf{k}}}\frac{-i\kappa u^*_{{\emph{\textbf{k}}}}v_{{\emph{\textbf{k}}}}^*v_{{\emph{\textbf{k}}}}^2}{2E_{{\emph{\textbf{k}}}}}\vert l_0\rangle\bar{\gamma}_{{\emph{\textbf{k}}}\uparrow}\bar{\gamma}_{-{\emph{\textbf{k}}}\downarrow}\vert r_0\rangle+\sum_{\emph{\textbf{k}}}\frac{i\kappa u^{*2}_{\emph{\textbf{k}}}u_{\emph{\textbf{k}}}v_{\emph{\textbf{k}}}}{2E_{\emph{\textbf{k}}}^*}\gamma^{\dagger}_{{\emph{\textbf{k}}}\uparrow}\gamma^{\dagger}_{-{\emph{\textbf{k}}}\downarrow}\vert l_0\rangle\vert r_0\rangle-\sum_{\emph{\textbf{k}}}\frac{\kappa\vert v_{\emph{\textbf{k}}}\vert^4}{4\text{Im}(E_{\emph{\textbf{k}}})}\gamma^{\dagger}_{{\emph{\textbf{k}}}\uparrow}\gamma^{\dagger}_{-{\emph{\textbf{k}}}\downarrow}\vert l_0\rangle\bar{\gamma}_{{{\emph{\textbf{k}}}\uparrow}}\bar{\gamma}_{{-{\emph{\textbf{k}}}\downarrow}}\vert r_0\rangle.
\end{aligned}
\end{equation}
\end{widetext}

 The normalization condition is $\vert\mathcal{N}_0\vert^2(\langle l^{(0)}_0\vert+\langle l^{(1)}_0\vert)(\vert r^{(0)}_0
\rangle+\vert r^{(1)}_0\rangle)=1$, where $\mathcal{N}_0$ is the normalization constant of the right ground state.
Simple algebra yields $\vert\mathcal{N}_0\vert^2=1/(\sum_{\emph{\textbf{k}}}\frac{\kappa^2\vert u_{\emph{\textbf{k}}}
\vert^4\vert v_{\emph{\textbf{k}}}\vert^4}{4E_{\emph{\textbf{k}}}^{*2}}+\sum_{\emph{\textbf{k}}}\frac{\kappa^2
\vert u_{\emph{\textbf{k}}}\vert^4\vert v_{\emph{\textbf{k}}}\vert^4}{4E_{\emph{\textbf{k}}}^{2}}+\sum_{\emph{\textbf{k}}}
\frac{\kappa^2\vert u_{\emph{\textbf{k}}}\vert^4\vert v_{\emph{\textbf{k}}}\vert^4}{16\text{Im}(E_{\emph{\textbf{k}}})^{2}}+1)$.
The first order corrections to the other eigenvectors
can be computed in the same way. These eigenvectors
form a new biorthonormal and complete basis, and  the whole dynamic can be predicted by Eq. (\ref{eq:sumrho}).
The calculations is tedious and  expression is involved, so we do not present them here.

\begin{figure}[htbp]
\centering
\includegraphics[width=0.48\textwidth]{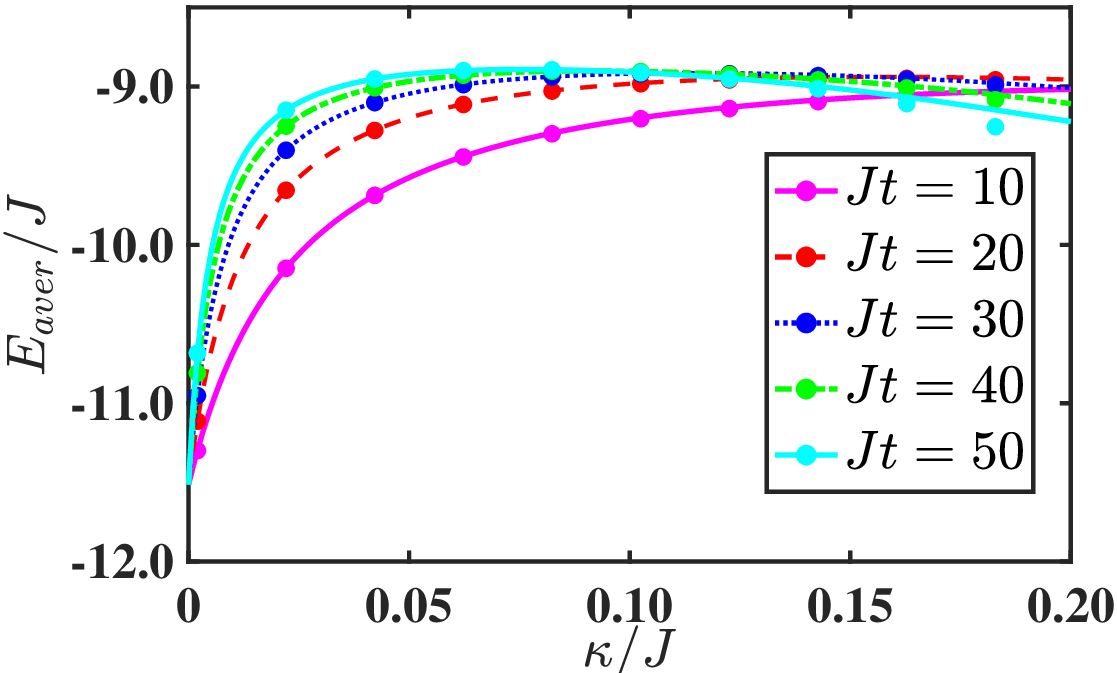}
\caption{The average energy $E_{aver}$ as a function of loss rate $\kappa$ at different time $t$. We calculate this energy by both the master equation (solid lines) and the perturbation theory (dotted lines).
The purple solid line, red dashed line, blue dotted line, green dash dotted line and cyan solid line are for different times,  $Jt=10,20,30,40,50$, respectively. Parameter $U_0=1.8J$  and the number of lattice sites is 10.}\label{Fig. sub.multitEk}
\end{figure}

 In Fig. \ref{Fig. sub.1}, we plot the average energy of the system at time $t$. The results are given by first order perturbation (blue solid line) and by numerical simulations with the master
equation (red circles). For the comparison purpose, the results given
 by non-Hermitian Hamiltonian (green solid line) are also shown. We can find
that the results given by perturbation theory match well with that by
the master equation even for long time evolution.
The other interesting observation is that
the energy has a large degree of deviation from the prediction based on the non-Hermitian Hamiltonian.
Remind that energy by non-Hermitian evolution is always unchanged because the initial state (i.e., the ground state) is a steady state of the system.
This feature can be understood by examining the fidelity of ground state $F=|\text{Tr}(\rho_0\rho)|/\sqrt{\text{Tr}(\rho_0^2)\text{Tr}(\rho^2)}$  \cite{wang2008alternative}.
As showed in Fig. \ref{Fig. sub.2}, at $Jt=100$ the system is almost all excited to
the excited states, resulting in a low fidelity between the initial states.
This result suggests that  the quantum jumps play an important role in this system, and the state $\vert r_0\rangle$ is unstable
under the effect of quantum jumps.

From the other point of view, non-Hermitian BCS model does not conserve the number of
fermions and thus the Liouvillians without and with quantum jumps can not be written
into a block-diagonal and block-upper-triangular form, respectively. This leads to
different spectrum for the Liouvillians with and without quantum jumps
\cite{yoshida2020fate,torres2014closedform}. This observation is quite
different from that of the non-Hermitian Hatano-Nelson model.
From the view point of quasiparticle,
the quasiparticle number conserves because $\left[H_{MF},\sum_{\emph{\textbf{k}}\sigma}\bar{\gamma}_{\emph{\textbf{k}}\sigma}
\gamma_{\emph{\textbf{k}}\sigma}\right]=0$.
However, in the basis spanned by the quasiparticles,  the jumping terms $\sum_{\emph{\textbf{k}}}\kappa c_{-\emph{\textbf{k}}\downarrow}
c_{\emph{\textbf{k}}\uparrow}$ still can not be written as block-upper-triangular form, which is in agreement with the aforementioned analysis.

In Fig. \ref{Fig. sub.multitEk}, we perform a comparison between the results of
the master equation and by the first order perturbation with different
loss rate $\kappa$. We observe that the bigger the loss rate is, the more intensity the excitation will be at the beginning.
Moreover, even when $\kappa$ is small, after long time, the average energy of the system suffers abruptly change.
 As $\kappa$ increase, the whole dynamic is totally different from NH ones (The same property as the green solid in Fig. \ref{Fig. sub.1}).

\begin{figure}[htbp]
\hspace{-2mm}
\centering
\subfigure{
\label{Fig. sub.MEmultin}
\includegraphics[width=0.24\textwidth]{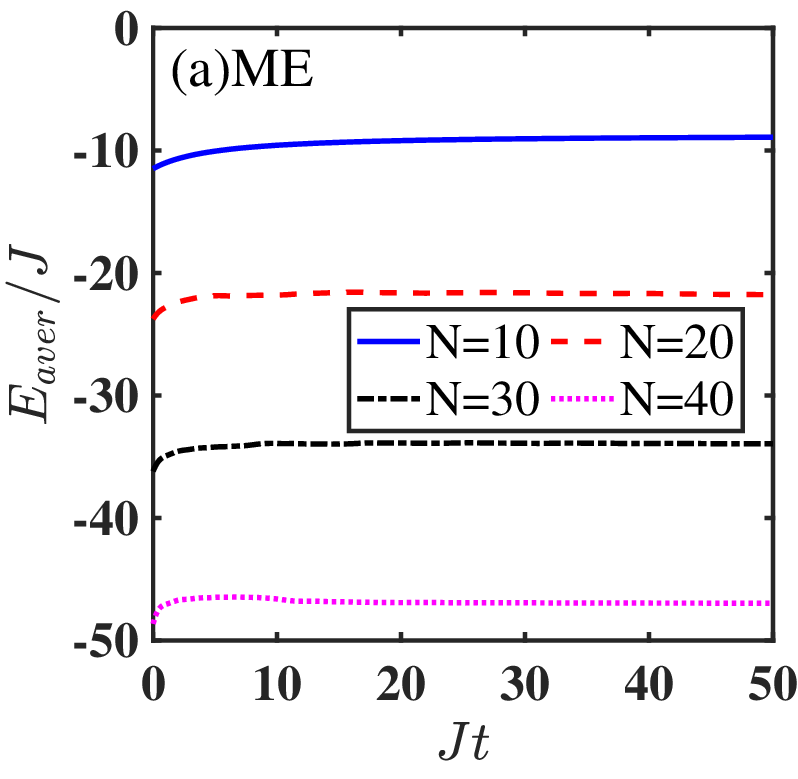}}\subfigure{
\label{Fig. sub.Permultin}
\includegraphics[width=0.24\textwidth]{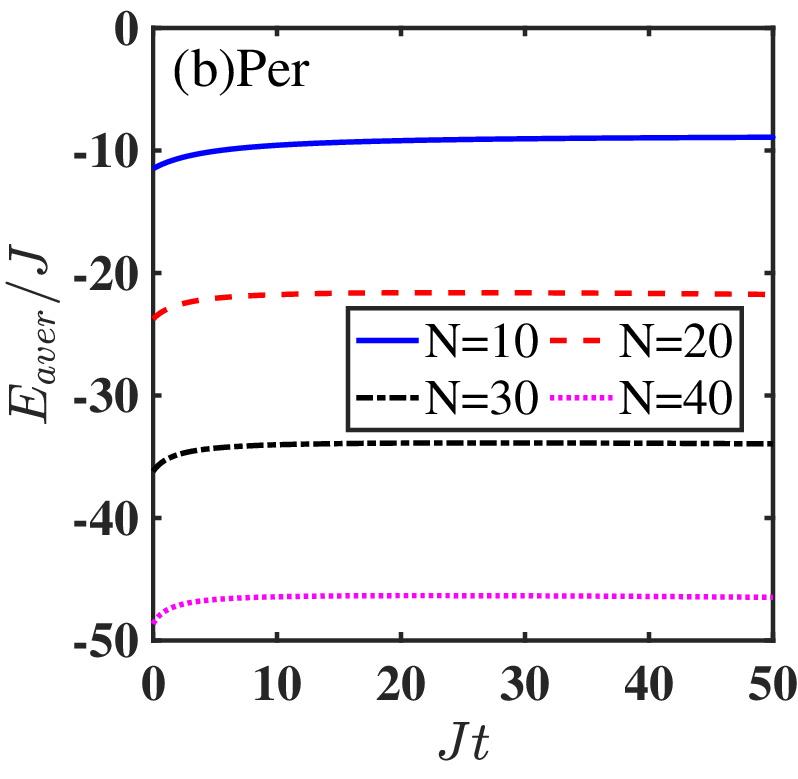}}
\caption{The time evolution of average energy $E_{aver}$  for  different $N$. $U_0=1.8J$ and the loss rate $\kappa=0.05J$ were set  for these plots.
 (a) are the results from the master equation, and (b) are the corresponding results from the
 perturbation theory. The blue solid line, red dashed line, black dash-dotted line and cyan dotted line
represent $N=10,20,30,40$ respectively.}
\end{figure}

Fig. \ref{Fig. sub.MEmultin} and  \ref{Fig. sub.Permultin} shows the time evolution  of the average energy with different system sizes. The results are calculated with
both the master equation and the  perturbation theory. The fact is that the
two results (one from the master
equation while another from the perturbation theory) matches well
even for system of big size suggest the validity of the our theory, which is also essential in many-body systems.

\begin{figure}[htbp]
\centering
\includegraphics[width=0.48\textwidth]{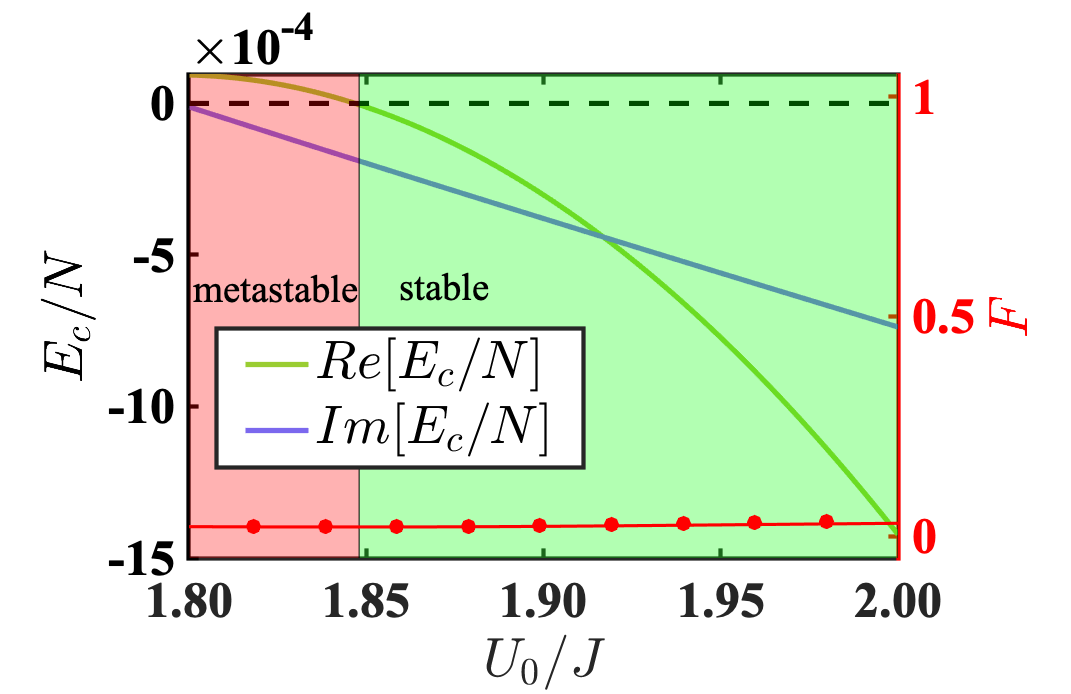}
\caption{Left axis: condensation energy $E_c$ as a function of $U_1/J$, $\kappa/J=0.1$, as $U_1/J$ increases, the superfluid undergoes a phase transition, from metastable area (red) to stable area (green). Right axis: fidelity of ground state at time $Jt=50$, the red solid line represent ME simulation, while the red dots stand for the perturbation, there is no significant difference in system dynamics.  }\label{Fig. EcF}
\end{figure}

Finally, we consider the condensation energy $E_c$ of the effective model that changes with the interaction strength $U_0$, with a fixed loss rate $\kappa=0.1J$, as shown in Fig. \ref{Fig. EcF}. As $U_0$ increases from $1.8J$ (as in Fig. \ref{Fig:Ec1}) to $2J$ (as in Fig. \ref{Fig:Ec2}), the real part of $E_c$ crosses the zero point, which leads to the transition of the superfluidity from the metastable (red area) to the stable region (green area). Similar result is also addressed in Ref. \cite{yamamoto2019theory}.
 However, at $Jt=50$, when quantum jumps are introduced, there is no significant variation in the fidelity $F$ from the ground state. It illustrates that in the study of NH systems, the presence of quantum jump events may challenge certain properties of NH systems, which warrants further investigation. Our proposed method can serve as a solid foundation for such research and verification.

\section{\label{sec:conclusion}CONCLUSION}

Based on the effective Hamiltonian approach, we  developed a perturbation theory to
study open quantum systems governed by the Lindblad master equation. Treating the
quantum jumps as   perturbations, we derived a set of corrections up to the first and second order in the jumps to the eigenenergies and corresponding eigenfunctions. This development is not trivial since our
perturbation theory is based on the non-Hermitian Hamiltonian and then the basis of the Hilbert space
behaves differently from its Hermitian counterpart. We applied our theory to two examples,
a decoherence two-level system and the non-Hermitian BCS model. The results show that  the present
theory is in good agreement with the results obtained by solving the master equation.
Besides, the present theory saves the computing time and in most cases analytical expressions can be found. We believe that the present theory opens a door to study
the non-Hermitian physics and  pave a way to check the validity of the
description of open systems by non-Hermitian Hamiltonians.

\begin{acknowledgments}
This work was supported by National Natural Science
Foundation of China (NSFC) under Grants No. 12175033,
No. 12147206 and National Key R$\&$D Program of China (No. 2021YFE0193500)
\end{acknowledgments}

\appendix

\section{THE DERIVATION OF EFFECTIVE HAMILTONIAN APPROACH\label{app:effectH} }

In this appendix, we will introduce the effective Hamiltonian approach, which was proposed to exactly solve the
master equation of open quantum systems.

According to the master equation Eq. (\ref{eq:ME}), the differential equation  of matrix element  $\rho_{mn}=
\langle l_m\vert\rho\vert l_n\rangle$ can be written as
\begin{equation}
   i\dot{\rho}_{mn}=\sum_{mn}\langle l_m\vert(H\rho-\rho H^{\dagger}+i\kappa F\rho F^{\dagger})\vert l_n\rangle\label{eq:element},
\end{equation}
where $H=H_0-i\kappa/2F^{\dagger} F$ is the effective non-Hermitian Hamiltonian.

For simplicity we assume that there is only one Lindblad operator in the master equation Eq. (\ref{eq:ME}).
We introduce an auxiliary system A (the ancilla, which is the same as the system), and then extend the    $N$-dimensional Hilbert space
to an $N^2$-dimensional Hilbert space. The Hilbert space of  the composite system (the open quantum system plus
the ancilla) can be expanded by a set of biorthonormal basis $\{|l_m\rangle\otimes|L_n\rangle^{A*},\,
|r_m\rangle\otimes|R_n\rangle^{A*}\}$, where $|R_n\rangle^A$ and $|L_n\rangle^A$ are the right and left basis for the ancilla Hilbert space.
At this time, a density matrix $\rho$ of the open quantum system can be mapped into a pure bipartite state in
$N^2$-dimensional Hilbert space, i.e.,
 \begin{equation}
\vert\psi_{\rho}(t)\rangle=\sum^N_{mn}\rho_{mn}(t)\vert r_m\rangle\vert R_n\rangle^{A*}.\label{eq:maprule}
\end{equation}

Taking the time derivative on the pure state $\vert\psi_{\rho}(t)\rangle$, it yields
 \begin{equation}
i\partial_t\vert\psi_{\rho}(t)\rangle=\sum^N_{mn}\left(\langle l_m\vert(H\rho-\rho H^{\dagger}
+i\kappa F\rho F^{\dagger})\vert l_n\rangle\right)\vert r_m\rangle\vert R_n\rangle^{A*}.\label{psit}
\end{equation}

For an arbitrary system operator $O$, we can define a corresponding operator $O^A$ of the ancilla, which satisfies
$\langle r_m\vert O^\dagger\vert l_n\rangle=(^A\langle L_n\vert O^A\vert R_m\rangle^A)^*$. Inserting the complete
relation Eq. (\ref{cr}) into Eq. (\ref{psit}), it is not difficult to obtain
 \begin{eqnarray}
\sum_{mn}\langle l_m\vert H\rho\vert l_n\rangle\vert r_m\rangle\vert R_n\rangle^{A*}&=&H\otimes I^A|\psi_\rho(t)\rangle,\nonumber\\
\sum_{mn}\langle l_m\vert\rho H^\dagger\vert l_n\rangle\vert r_m\rangle\vert R_n\rangle^{A*}&=&I\otimes (H^{A})^*|\psi_\rho(t)\rangle,\nonumber\\
 \sum_{mn}\langle l_m\vert F\rho F^\dagger\vert l_n\rangle\vert r_m\rangle\vert R_n\rangle^{A*}&=& F\otimes(F^{A})^*\vert\psi_{\rho}(t)\rangle.\nonumber
\end{eqnarray}

Thus, the dynamical equation for the composite system can be rewritten as
\begin{eqnarray}
i\partial_t\vert{\psi}_{\rho}(t)\rangle=\widetilde{H}\vert\psi_{\rho}(t)\rangle, \label{eq:psidt}
\end{eqnarray}
with the effective Hamiltonian $\widetilde{H}=H\otimes I-I\otimes(H^{A})^*+i\kappa F\otimes (F^{A})^*.$
Therefore, the master equation is equivalent to the evolution of a pure state of the composite system. The jumping terms in the master equation
describe the interactions between the open quantum system and the ancilla. More, if we choose the basis of the ancilla the same representation as the  system,
the effective Hamiltonian returns to the Liouvillian superoperator ( a $-i$ factor is neglected) \cite{minganti2019quantum},
\begin{equation}
    \widetilde{H}=(H_0-\frac{i\kappa F^{\dagger}F}{2})\otimes I-I\otimes(H_0^{\text{TR}}+\frac{i\kappa F^{\text{TR}}F^*}{2})+i\kappa F\otimes F^*,
\end{equation}
where TR denotes transpose operation.

\section{A SHORT DERIVATION OF THE EQUIVALENCE OF BIORTHOGONAL BASIS\label{app:Vmatrix} }

Consider an arbitrary non-Hermitian system with a Hamiltonian $H_{NH}$ which is diagonalizable and nondegenerate.
The biorthonormal basis $\{\vert r_n\rangle\},\{\vert l_n\rangle\}$ are the right and left eigenvectors of $H_{NH}$, which
satisfy Eq. (\ref{eq:bio}). In the following, we divide our \text{discussion} into two cases, i.e., the real spectrum and the complex
spectrum \cite{wang1979disscussion}.

Now we first prove that $H_{NH}$ with a real spectrum $\{E_n\}$ can be written as $H_{NH}=(A^{-1})^{\dagger}H_{H}
A^{\dagger}$, where $H_{H}$ is a Hermitian Hamiltonian and $A$ is a non-unitary operator.According to the completeness
relation $\langle l_n\vert r_m\rangle=\delta_{mn}$, the left eigenvector $|l_n\rangle$ can be expressed as an expansion of
the right eigenvetors, i.e.,
\begin{equation}
\vert l_n\rangle=\sum_{n'}M_{nn'}\vert r_{n^\prime}\rangle,\label{eq:lr}
\end{equation}
 taking an Hermitian conjugate operation on above equation and acting $|r_{n'}\rangle$ on the result from the right hand side,
it yields
\begin{equation}
 \langle l_n\vert r_{n^\prime}\rangle=\sum_{n''}M^*_{nn''}\langle r_{n''}\vert r_{n^\prime}\rangle=\delta_{nn^\prime}.\label{eq:M}
 \end{equation}

From Eq. (\ref{eq:M}), it is easy to obtain $(M^*)X=I$, where $X_{mn}=\langle r_{m}\vert r_{n}\rangle$ and $I$ is the identity matrix. Thus
\begin{align}
(M^{*-1})_{mn}=\langle r_{m}\vert r_{n}\rangle, \label{eq:M1}\\
(M^{-1})_{mn}=\langle r_{n}\vert r_{m}\rangle.\label{eq:M2}
\end{align}

Combing Eq. (\ref{eq:M1}) and (\ref{eq:M2}), it is straightforward to find that $(M^{*-1})_{nm}=(M^{-1})_{mn}$ always holds for arbitrary $m,n$. Recall the definition of the Hermitian matrix, $M^{-1}$ is a Hermitian matrix, which implies that $M$ is Hermitian. In other words, the left
eigenvectors $\{\vert l_n\rangle\}$ can be obtained from the right eigenvectors $\{\vert r_n\rangle\}$ via a Hermitian transformation.
Considering an arbitrary set of complete orthonormal basis $\{\vert{n}\rangle\}$, we have
$(M^{-1})_{nm}=\langle r_n\vert r_m\rangle=\sum_{n^\prime}\langle r_n\vert{n^\prime}\rangle\langle{n^\prime}\vert r_m
\rangle=(A^{\dagger-1}A^{-1})_{nm}\!$. By defining $(A^{\dagger-1})_{nn^\prime}=\langle r_n\vert{n^\prime}\rangle$, it can be seen that
\begin{equation}
U=AA^\dagger,\label{eq:U}
 \end{equation}
 as a result,  we obtain Eq. (\ref{eq:Ainvertible}) from Eq. (\ref{eq:U}) and Eq. (\ref{eq:lr}).

For a non-Hermitian $H_{NH}$ with a complex energy spectrum $\{E_n\}$, we can always find two Hermitian Hamiltonian $H_1=H_1^{\dagger}$,
$H_2=H_2^{\dagger}$ which satisfy $H_{NH}=H_1+iH_2$. We still have two cases such as $H_{NH1}:\left[H_1,H_2\right]\!=\!0$ and
$H_{NH2}:\left[H_1,H_2\right]\!\neq\! 0$ . For the former situation, despite the $H_{NH}$ is non-Hermitian, $H_1$ and $H_2$ share common
 orthonormal basis $\{\vert n\rangle\}$. For the latter case, it can be proved that $H_{NH2}$ is of an equivalence relation with $H_{NH1}$. According to
  Eq. (\ref{eq:U}), we have $\vert l_n\rangle=U\vert r_n\rangle=AA^{\dagger}\vert r_n\rangle$ and $A^{-1}\vert l_n\rangle=A^{\dagger}\vert r_n
  \rangle=\vert n\rangle$. $\{\vert n\rangle\}$ is a set of orthonormal basis which relate to a non-Hermitian Hamiltonian $H_{NH1}$.
  Thus, it yields $A^{\dagger}H_{NH2}(A^\dagger)^{-1}\vert n\rangle=E_n\vert n\rangle$ and $A^{-1}H_{NH2}A\vert n\rangle=
  E_n^*\vert n\rangle$. Therefore, Eq. (\ref{eq:Ainvertible}) always holds.
\\

\nocite{*}
\bibliographystyle{unsrt}

\end{document}